\documentclass[prl,nofootinbib,twocolumn]{revtex4}
\def\mysection#1{{\bf #1.} }
\def\mysections#1{{\bf #1.} }

\usepackage{amssymb}
\usepackage{amsmath}
\usepackage[dvips]{graphicx}
\usepackage{longtable}
\usepackage{verbatim}
\usepackage{amsfonts}

\newcommand{\be}{\begin{equation}}
\newcommand{\ee}{\end{equation}}
\newcommand{\bea}{\begin{eqnarray}}
\newcommand{\eea}{\end{eqnarray}}
\newcommand{\beq}{\begin{equation}}
\newcommand{\eeq}{\end{equation}}
\def\beqa{\begin{eqnarray}}
  \def\eeqa{\end{eqnarray}}
\newcommand{\bv}{\left(\begin{array}{c}}
\newcommand{\ev}{\end{array}\right)}
\newcommand{\no}{\nonumber}
\def\lsim{\mathrel{\rlap{\lower4pt\hbox{\hskip1pt$\sim$}}
    \raise1pt\hbox{$<$}}}         %less than or approx. symbol
\def\gsim{\mathrel{\rlap{\lower4pt\hbox{\hskip1pt$\sim$}}
    \raise1pt\hbox{$>$}}}         %greater than or approx. symbol

\begin{document}

\vspace*{-30mm}

\title{\boldmath Relating direct CP violation in $D$ decays\\ and the
  forward-backward asymmetry in $t\bar t$ production}

\author{Yonit Hochberg}\email{yonit.hochberg@weizmann.ac.il}
\affiliation{Department of Particle Physics and Astrophysics,
  Weizmann Institute of Science, Rehovot 76100, Israel}

\author{Yosef Nir}\email{yosef.nir@weizmann.ac.il}
\affiliation{Department of Particle Physics and Astrophysics,
  Weizmann Institute of Science, Rehovot 76100, Israel}

\vspace*{1cm}

\begin{abstract}
  The CDF and LHCb experiments have recently provided two intriguing
  hints for new physics: a large forward-backward asymmetry in $t\bar
  t$ production, and a direct CP asymmetry in $D$ decays of order a
  percent. In both cases, flavor non-universal interactions are
  required in the up sector, raising the possibility that the two
  effects come from one and the same new physics source. We show that
  a minimal model, with an extra scalar doublet, previously suggested to
  explain the top data, gives~-- without any modifications or
  additions -- a contribution to CP violation in charm decays that is
  of the right size.
\end{abstract}

\maketitle

%%%%%%%%%%%%%%%%%%%%
\mysection{Introduction to $\Delta A_{CP}$}
The LHCb experiment has announced evidence for direct CP violation in
singly Cabibbo suppressed $D$ decays \cite{arXiv:1112.0938},
\beqa\label{eq:acp}
\Delta A_{CP}&\equiv&
A_{CP}(K^+K^-)-A_{CP}(\pi^+\pi^-)\no\\
&=&(-0.82\pm0.21\pm0.11)\times10^{-2}.
\eeqa
The updated world average for this asymmetry is~\cite{arXiv:1111.4987}
$\Delta A_{CP}=(-0.65\pm0.18)\times10^{-2}$, which is more than
$3.5\sigma$ away from zero. Here,
\beq
A_{CP}(f)=\frac{\Gamma(D^0\to f)-\Gamma(\overline{D}^0\to f)}
{\Gamma(D^0\to f)+\Gamma(\overline{D}^0\to f)}.
\eeq
In $\Delta A_{CP}$, that is the difference between asymmetries,
effects of indirect CP violation cancel out
\cite{Grossman:2006jg}. (Due to different decay time acceptances
between the $K^+K^-$ and $\pi^+\pi^-$ modes, a small residual effect
of indirect CP violation remains.) Thus, $\Delta A_{CP}$ is a
manifestation of CP violation in decay.

The Standard Model (SM) contribution to the individual asymmetries is
CKM-suppressed by a factor of
\beq\label{eq:ckmsup}
I_{\rm CKM}\equiv2{\cal
  I}m\left(\frac{V_{ub}V_{cb}^*}{V_{us}V_{cs}^*}\right)
\approx1.2\times10^{-3},
\eeq
and loop-suppressed by a factor of order $\alpha_s(m_c)/\pi\sim0.1$.
(For the numerical estimate of Eq. (\ref{eq:ckmsup}) we use \cite{pdg}
$|V_{cb}|=0.041$, $|V_{ub}|=0.0035$, $|V_{us}|=0.23$, $|V_{cs}|=0.97$,
and $\sin\gamma=0.93$.)  While perhaps one cannot exclude an
enhancement factor of order 30 from hadronic physics
\cite{Brod:2011re}, in which case (\ref{eq:acp}) will be accounted for
by SM physics, this situation seems unlikely. It is thus interesting
to find new physics that can contribute to $\Delta A_{CP}$ a factor of
order ten higher than the SM
\cite{Grossman:2006jg,arXiv:1111.4987,Rozanov:2011gj}.

%%%%%%%%%%%%%%%%%%%%
\mysection{Introduction to $A_{FB}^{t\bar t}$}
The CDF collaboration has announced evidence for a large
forward-backward $t\bar t$ production asymmetry for large invariant
mass of the $t\bar t$ system~\cite{Aaltonen:2011kc}:
\beq\label{eq:atthexp}
A^{t\bar t}_h\equiv A^{t\bar t}_{FB}(M_{t\bar t}\geq450\ {\rm
  GeV})=+0.475\pm0.114\,,
\eeq
to be compared with the SM prediction~\cite{Almeida:2008ug,
  Bowen:2005ap, Antunano:2007da}, $\left(A^{t\bar
    t}_h\right)_{SM}=+0.09\pm0.01$. Eq.~(\ref{eq:atthexp}) updates
(and is consistent with) previous CDF and D0 measurements of the
inclusive asymmetry~\cite{:2007qb,Aaltonen:2008hc}.

The source of the asymmetry must be in the quark process $u\bar u\to
t\bar t$. The large effect is suggestive of interference between a
tree level exchange of a new boson with an electroweak-scale mass and
the SM gluon-mediated amplitude (see \cite{Kamenik:2011wt} and
references therein). Moreover, the couplings of the intermediate boson
cannot be flavor universal.

It is interesting to note that both $\Delta A_{CP}$ and $A^{t\bar
  t}_{FB}$ are related to flavor physics in the up sector. Could the
two measurements be related to each other? In this work, we show
that a mechanism previously studied to explain
$A_{FB}^{t\bar t}$ \cite{arXiv:1107.4350} {\it predicts} a new
physics contribution to $\Delta A_{CP}$ that is quantitatively of the
right size, namely a factor of ${\cal O}(10-100)$ above the SM.

%%%%%%%%%%%%%%%%%%%%
\mysection{Scalar mediated $A_{FB}^{t\bar t}$}
In Ref. \cite{arXiv:1107.4350} we investigated (in collaboration with
K. Blum) whether the large value reported by CDF for $A_{FB}^{t\bar
  t}$ at large invariant mass $M_{t\bar t}$ can be accounted for by
tree level scalar exchange.  We considered top-related measurements,
flavor constraints, and electroweak precision measurements. We reached
the following conclusions:
\begin{itemize}
\item Out of the eight possible scalar representations that are
  relevant to $A_{FB}^{t\bar t}$, only the color-singlet weak-doublet,
\beq
\Phi\sim(1,2)_{-{1}/{2}}=\bv\phi^0\\ \phi^-\ev,
\eeq
can enhance $A_h^{t\bar t}$ and remain consistent with the low bin
$t\bar t$ asymmetry and the total and differential $t\bar t$ cross
section.  Roughly speaking, the relevant Yukawa coupling should be
$\mathcal{O}(1)$, and the mass of the scalar should be below $\sim
130$~GeV. (See also
\cite{Nelson:2011us,AguilarSaavedra:2011vw,Babu:2011yw,Grinstein:2011dz}.)
\item Two types of couplings of $\Phi$ can contribute to $u\bar u \to
  t\bar t$: $X_{13} q^\dagger_{L1} \Phi t_R$ and $X_{31}
  q^\dagger_{L3} \Phi u_R$. There is no tension with the differential
  or total $t\bar t$ production cross section. Both couplings are
  constrained by flavor physics:
\begin{enumerate}
\item The $X_{13}$ coupling is strongly constrained by
  $K^0-\overline{K}{}^0$ and/or $D^0-\overline{D}{}^0$ mixing, and so
  cannot generate a large $A^{t\bar t}_h$.
\item The $X_{31}$ coupling is not strongly constrained by neutral
  meson mixing, or by $R_b$. If $\phi^-$ couples to the three
  left-handed down generations with CKM-like suppression ${\cal
    O}(V_{tq})$, then it contributes to the branching ratio of
  $\overline{B^0}\rightarrow\pi^+ K^-$ more than two orders of
  magnitude above the experimental bounds. If, on the other hand, the
  $X_{31}$ coupling is carefully aligned so that $\phi^-$ couples only
  to $b_L$ (but not to $s_L$ and $d_L$), then it can be large enough
  to explain $A^{t\bar t}_h$.
\end{enumerate}
\end{itemize}
Thus, the relevant Lagrangian terms for the new weak doublet field
are given in the quark mass basis as follows \cite{arXiv:1107.4350}:
\beq\label{eq:defx}
{\cal L}_\Phi=-V(\Phi)+2\lambda\left[
  \phi^0 (U_{L})_i^\dag V_{ib}u_{R}+2\phi^-b_{L}^\dag u_{R}+{\rm
    h.c.}\right],
\eeq
where $(U_L)_{1,2,3}=u_L,c_L,t_L$. The $\lambda V_{tb}\phi^0 \bar t_L
u_R$ coupling accounts for the forward-backward asymmetry in $t\bar t$
production, with
\beq\label{eq:afbcstrt}
|\lambda|\gsim0.6,\;\;\;M_\Phi\lsim130\,{\rm GeV}.
\eeq
(For further details see Ref.~\cite{arXiv:1107.4350}.)

%%%%%%%%%%%%%%%%%%%%
\mysection{Scalar mediated $\Delta A_{CP}$}
In addition to the coupling to $\bar t_L u_R$, the neutral scalar
$\phi^0$ couples $u_R$ to the lighter two up-type quarks:
$\lambda V_{cb}\phi^0 \bar c_L u_R+\lambda V_{ub}\phi^0 \bar u_L
u_R$. Integrating out the $\phi^0$ field, these couplings lead to the
following effective four-quark coupling:
\beq\label{eq:cuuu}
\frac{4|\lambda|^2}{m_{\phi^0}^2}V_{ub}V_{cb}^*(\bar u_R c_L)(\bar u_L
u_R).
\eeq
This operator contributes, via annihilation diagram ($c\bar u\to u\bar
u$), to both $D^0\to K^+K^-$ and  $D^0\to\pi^+\pi^-$ decays. In the
U-spin symmetry limit, the resulting asymmetries in the two modes are
equal in magnitude and opposite in sign.

Thus, the expected size of $\Delta A_{CP}$ from the interference
between the new physics (\ref{eq:cuuu}) amplitude and the SM
$W$-mediated tree amplitude is
\beqa\label{eq:dacpphi}
\Delta A_{CP}&=&2\sqrt{2}(G_0/G_F)I_{\rm CKM}I_{\rm QCD}\no\\
&\sim&(2-7)\times10^{-2}I_{\rm QCD}.
\eeqa
The various factors in this equations are the following:
\begin{itemize}
\item The factor of 2 comes from the opposite sign asymmetries in the
  U-spin limit.
\item $G_0$ is defined as $G_0\equiv4|\lambda|^2/m_{\phi^0}^2$.
  Eq.~(\ref{eq:afbcstrt}) implies that $G_0/(G_F/\sqrt{2})\sim10-30$.
\item $I_{\rm CKM}$ is the CKM suppression factor defined in
  Eq. (\ref{eq:ckmsup}). Its value is known to a good approximation,
  including the CP violating phase.
\item $I_{\rm QCD}$ includes all the hadronic aspects of the decay:
  ratio of matrix elements, the price for annihilation (if any),
  U-spin violation, and the strong phase.
\end{itemize}
Thus, on one hand, all the electroweak parameters are well known but,
on the other hand, the hadronic physics introduces an order of
magnitude uncertainty.

Compared to the SM, the scalar contribution is tree level, and a loop
suppression, naively of order $\alpha_s(m_c)/\pi\sim0.1$, is
avoided. Moreover, the contribution is enhanced by the requirement
that $G_0\gg G_F$ (to account for $A_{FB}^{t\bar t}$).
It involves, however, annihilation, which introduces a suppression
factor that is naively of order $f_D/m_D\sim0.1$. Ref.
\cite{Brod:2011re} argues, based on experimental data, that tree level
annihilation amplitudes are large, and do not suffer $1/m_c$
suppression. In any case, it is plausible that hadronic physics, {\it
  e.g.} the strong phase, provides the mild suppression, $I_{\rm
  QCD}\sim0.1-0.3$, that is necessary to make the theoretical
prediction (\ref{eq:dacpphi}) consistent with the experimental result
(\ref{eq:acp}).

We conclude that our model predicts $\Delta A_{CP}$ of order a
percent.

%%%%%%%%%%%%%%%%%%%%
\mysection{Scalar mediated $\epsilon^\prime/\epsilon$}
The same Yukawa couplings of $\phi^0$ that contribute unavoidably to
direct CP violation in $D$ decays, contribute unavoidably also to
direct CP violation in $K$ decays. The former effect comes at tree
level and modifies $\Delta A_{CP}$. The latter effect comes via a box
diagram, involving $\phi^0$ and a $W$-boson, and modifies
$\epsilon^\prime/\epsilon$. This type of relation was pointed out in
Ref.~\cite{arXiv:1111.4987}. Their analysis cannot, however, be
directly applied to our model, since it makes use of an effective
Lagrangian with a scale of new physics that is similar to or higher
than 1 TeV. Instead, we carried out a full calculation of the relevant
box diagram.

We find that the couplings of Eq. (\ref{eq:defx}) lead to the
following effective four-quark coupling:
\beq\label{eq:epsp}
\frac{\sqrt 2 |\lambda|^2 G_F}{\pi^2}\
\frac{\ln x_\phi}{1-x_\phi}\
V_{ud}^*V_{cs}V_{ub}V_{cb}^*(\bar d_L u_R)(\bar u_R s_L),
\eeq
where $x_\phi\equiv m_{\phi^0}^2/m_W^2$. To estimate the contribution
of this operator to $\epsilon^\prime/\epsilon$, we use
\beq\label{eq:epspgen}
{\cal R}e\left(\frac{\epsilon'}{\epsilon}\right)=
-\frac{\omega}{\sqrt 2|\epsilon|{\cal R}eA_0}
\left({\cal I}mA_0-\frac{1}{\omega}{\cal I}mA_2\right).
\eeq
We further use the recent lattice calculation of the relevant matrix
element \cite{Blum:2011ng}, and obtain, for the scalar mediated
contribution,  $[{\cal I}m A_2]^\phi=-(5.6-7.7)\times10^{-12}$ GeV, a
factor of $8-11$ above the SM value \cite{Blum:2011ng}, $[{\cal I}m
A_2]^{\rm SM}=-(6.8\pm1.4)\times10^{-13}$ GeV.  Inserting these ranges
into Eq.~(\ref{eq:epspgen}), we find ${\cal
  R}e(\epsilon^\prime/\epsilon)_\phi=-(6.3\pm2.3)\times10^{-3}$,
compared to \cite{Blum:2011ng} ${\cal
  R}e(\epsilon^\prime/\epsilon)_{\rm EWP}=-(6.5\pm1.3)\times10^{-4}$.

Our result for ${\cal I}m A_2$ can be combined with the experimental
results for ${\cal R}e A_2$, ${\cal R}e A_0$, and
$\epsilon^\prime/\epsilon$ to obtain the unknown ratio:
\beq
\frac{{\cal I}m A_0}{{\cal R}e A_0}=-(4-7)\times10^{-4},
\eeq
which is a factor of about 3 above the value extracted within the
Standard Model \cite{Blum:2011ng}, $\frac{{\cal I}m A_0}{{\cal R}e
  A_0}=-(1.6\pm0.3)\times10^{-4}$.  Given the large hadronic
uncertainties \cite{Blum:2011pu}, such an enhancement cannot be used
to exclude our model. Below we mention other ways in which the model
can be tested.  We note, however, that had it been possible to exclude
the model based on $\epsilon^\prime/\epsilon$, it would have led to
the interesting result that there is no viable single scalar-mediated
mechanism that can explain the large value of $A_h^{t\bar t}$.

%%%%%%%%%%%%%%%%%%%%
\mysection{Additional phenomenological aspects}
In this section we assume throughout that Eq.~\eqref{eq:defx}
describes the full set of interactions of the scalar weak-doublet with
fermions, and that $\phi^0$ is a mass eigenstate. We postpone the
discussion of additional couplings, beyond those that are required to
explain $A_{FB}^{t\bar t}$, to future work.

The scalar exchange contributes to $D^0-\overline{D}{}^0$ mixing via box
diagrams. Requiring that this contribution is not larger than the
experimental constraint from $\Delta m_D$ gives \cite{arXiv:1107.4350}
\beq
\frac{|\lambda|^4}{32\pi^2}\left(\frac{100\ {\rm
      GeV}}{m_{\phi^0}}\right)^2
(V_{ub}V_{cb}^*)^2<7\times10^{-9}.
\eeq
Given that $|\lambda|={\cal O}(1)$ and $m_{\phi^0}\approx100$ GeV, the
new contribution is a factor of order 100 below the experimental
value, which is negligibly small for both $\Delta m_D$ and indirect CP
violation \cite{Gedalia:2009kh}.

As concerns $D$ decays, the operator (\ref{eq:cuuu}) contributes to
neither Cabibbo favored, nor doubly Cabibbo suppressed decays. Thus it
affects only the singly Cabibbo suppressed decays. Given that it is
suppressed by the fifth power of the Cabibbo angle, the effects on the
rates of these decays is negligible, and it can be signalled only via
CP violation.

We note that this model predicts a contribution to the CP
asymmetry in both the $D^0\to K^+K^-$ and $D^0~\to~\pi^+\pi^-$ channels,
and so measurements of the individual asymmetries by the LHCb
collaboration would be useful. Previous experiments and, in
particular, the recent CDF result \cite{Aaltonen:2011se}, lead to a
world average of \cite{Asner:2010qj}
\beqa
A_{CP}(K^+K-)&=&-0.0023\pm0.0017,\no\\
A_{CP}(\pi^+\pi^-)&=&+0.0020\pm0.0022,
\eeqa
consistent with the U-spin prediction of equal magnitudes and opposite
signs.

The scalar exchange also contributes in principle to the LHC charge
asymmetry in top pair production. We find that the parameter space of
mass and coupling of the weak-doublet relevant for explaining the
$t\bar t$ forward-backward asymmetry and $\Delta A_{CP}$ is at present
unconstrained by the CMS~\cite{:2011hk} and ATLAS~\cite{ATLASac} results
at the $2\sigma$ level.

Since the required mass range for $m_{\phi^0}$ is $100-130$ GeV, the
question arises whether $\phi^0$ can be discovered via present Higgs
searches at ATLAS and CMS. Here the answer is, unfortunately,
negative. The reason is that the leading two-body decay mode of
$\phi^0$ is $\phi^0\to c\bar u$. The decay to $\gamma\gamma$ is
generated only by an up-quark loop and is suppressed by a $|V_{ub}|$
factor. Furthermore, $\phi^0$ does not couple to $W^+W^-$, $ZZ$, and
$b\bar b$ and $\tau^+\tau^-$. Thus, none of the decay modes that are
used in the search of the Higgs boson are useful to observe $\phi^0$.
The leading three-body decay mode of $\phi^0$ is $\phi^0 \to u\bar bW$
via an off-shell top.

The coupling of the charged scalar $\phi^+$ in Eq. (\ref{eq:defx}) is
to only $\bar bu$ pair. Therefore, it does not contribute to $B$
decays. It could be that $\Phi$ has couplings additional to those of
Eq.~(\ref{eq:defx}). For example, if it couples to $\tau\nu$, then it
can affect the $B\to\tau\nu$ decay. A discussion of additional
couplings beyond those of Eq.~(\ref{eq:defx}) is postponed to future
work.

Finally, as discussed in Ref.~\cite{arXiv:1107.4350}, the model
predicts a large cross section for single top production, but a
dedicated study is required in order to establish the applicability of
existing measurements.

%%%%%%%%%%%%%%%%%%%%
\mysection{Conclusions}
Evidence for a large forward-backward asymmetry in $t\bar t$
production ($A_{FB}^{t\bar t}$) has been observed by the CDF
collaboration.  Evidence for direct CP violation in singly Cabibbo
suppressed $D$ decays ($\Delta A_{CP}$) has been observed by the LHCb
collaboration. Both effects are suggestive of new physics that has
non-universal interactions in the up sector.

In previous work \cite{arXiv:1107.4350} it was found that, among the
single scalar mediated mechanisms that can explain $A_{FB}^{t\bar t}$,
only the $t-$channel exchange of a weak doublet, with a very special
flavor structure, is consistent with the total and differential $t\bar
t$ cross section, flavor constraints and electroweak precision
measurements. In this work we showed that the required flavor
structure implies that the scalar {\it unavoidably} contributes at
tree level to $\Delta A_{CP}$.  The relevant electroweak parameters
are either directly measured, or fixed by the top-related data,
implying that, for a plausible range of the hadronic parameters, the
scalar mediated contribution is of the right size.

The model predicts large effects on $\epsilon^\prime/\epsilon$ and on
single top production. It can be excluded based on better knowledge of
the hadronic parameters in the calculation of
$\epsilon^\prime/\epsilon$, or with a dedicated study of single top
production that takes into account the special features of the
model~\cite{arXiv:1107.4350}.

We find it intriguing that a single, highly constrained, mechanism
simultaneously explains the two measurements. It motivates further
study of possible experimental signatures and tests, as well as a
search for a theoretical framework that would give a natural
explanation to the required flavor structure.

%%%%%%%%%%%%
\mysections{Acknowledgments} We thank Elaine Goode, Alex Kagan and
Chris Sachrajda for numerous discussions and, in particular, for their
help in the analysis of $\epsilon^\prime/\epsilon$. We thank Gino
Isidori, Gilad Perez and Tomer Volansky
for useful discussions. YN is the Amos de-Shalit chair of theoretical
physics and supported by the Israel Science Foundation, and the
German-Israeli foundation for scientific research and development
(GIF).

%\vspace*{-5mm}
%%%%%%%%%%%%%%%%%%%%%

\end{document}